\documentclass[twocolumn,secnumarabic,amssymb, nobibnotes, aps, pre]{revtex4-2}
\usepackage{amsfonts,amssymb,amsmath} 
\usepackage{color} 
\usepackage{graphicx} 
\usepackage{epstopdf,mathrsfs} 
\usepackage[colorlinks,linktocpage,linkcolor=blue]{hyperref}

\begin{document}

\title{Aging two-state process with L\'{e}vy walk and Brownian motion}

\author{Xudong Wang}
\author{Yao Chen}
\author{Weihua Deng}
\affiliation{School of Mathematics and Statistics, Gansu Key Laboratory
of Applied Mathematics and Complex Systems, Lanzhou University, Lanzhou 730000,
P.R. China}

\begin{abstract}
With the rich dynamics studies of single-state processes, the two-state processes attract more and more interests of people, since they are widely observed in complex system and have effective applications in diverse fields, say, foraging behavior of animals. This report builds the theoretical foundation of the process with two states: L\'{e}vy walk and Brownian motion, having been proved to be an efficient intermittent search process.
The sojourn time distributions in two states are both assumed to be heavy-tailed with exponents $\alpha_\pm\in(0,2)$. The dynamical behaviors of this two-state process are obtained through analyzing the ensemble-averaged and time-averaged mean squared displacements (MSDs) in weak and strong aging cases. It is discovered that the magnitude relationship of $\alpha_\pm$ decides the fraction of two states for long times, playing a crucial role in these MSDs. According to the generic expressions of MSDs, some inherent characteristics of the two-state process are detected. The effects of the fraction on these observables are detailedly presented in six different cases. The key of getting these results is to calculate the velocity correlation function of the two-state process, the techniques of which can be generalized to other multi-state processes.
\end{abstract}

\maketitle

Searching a target is a natural demand in the real world. At the same time,
many physical or biological problems can be regarded as the search processes, describing how a searcher finds a target located in an unknown position. At the macroscopic scale, it is exemplified as animals searching for food or a shelter \cite{Bell:1991}. At the microscopic scale, one can cite the localization by a protein of a specific DNA sequence or the active transport of vesicles in cells \cite{BenichouLoverdoMoreauVoituriez:2011}. In these examples, the search time is generally a limiting quantity which has to be optimized by choosing different search strategies. Intermittent search strategies have been proved to play a crucial role in optimizing the search time of randomly hidden targets \cite{BenichouCoppeyMoreauSuetVoituriez:2005,LomholtKorenMetzlerKlafter:2008}. This kind of search behavior could be extended to broader research domains such as the theory of stochastic processes \cite{BartumeusCatalanFulcoLyraViswanathan:2002}, applied mathematics \cite{Stone:1975}, and molecular biology \cite{CoppeyBenichouVoituriezMoreau:2004}; and it also motivates some new interesting research topics \cite{XuDeng:2018,*XuDeng:2018-2}.

For the intermittent search process, it switches between two phases --- local Brownian search phase and ballistic relocation phase (L\'{e}vy walk). The searcher displays a slow reactive motion in the first phase, during which the target can be detected. The latter fast phase aims at relocating into unvisited regions to reduce oversampling, during which the searcher is unable to detect the target.
In the situation of rare targets, it has been shown that the search process with L\'{e}vy distributed relocations significantly outperforms that with exponentially distributed relocation  \cite{LomholtKorenMetzlerKlafter:2008}. 
While the two-state process effectively models intermittent strategy, it is also observed in the transport of the neuronal messenger ribonucleoproteins delivered to their target synapses \cite{SongMoonJeonPark:2018}, where a type of L\'{e}vy walk process is interrupted by the emerging of rest. 
The rest period can be very long, characterized by power-law distribution without finite mean. This phenomenon becomes a striking feature of the RNA transport in neuronal systems.

The intermittent strategy has been verified to be optimum for searching targets in some specific macroscopic and microscopic situations. But generally it is hard to believe that the intermittent strategy is always the best one in all the foraging behaviors of animals and the intracellular transport in microscopic scale. A question naturally comes up: How about the field of its application? Based on this motivation, it is necessary to build a complete theoretical foundation for this kind of two-state processes for dealing with data observed in experiments. In this report, we consider the two-state process mentioned earlier (i.e., the standard L\'{e}vy walk and Brownian motion) and mainly investigate their statistical behaviors, such as ensemble-averaged mean square displacement (EAMSD) and time-averaged mean square displacement (TAMSD). In particular, we carefully examine the aging behaviors of the two-state process, while the aging continuous-time random walk (CTRW) \cite{BarkaiCheng:2003}, aging renewal theory \cite{SchulzBarkaiMetzler:2014} and aging ballistic L\'{e}vy walks \cite{MagdziarzZorawik:2017} have been fully discussed. Since the observation time might not be the beginning of a process in experiments, aging behavior should be paid some attention and it may display interesting phenomena in anomalous diffusion processes \cite{SokolovBlumenKlafter:2001,Allegrini--etal:2002}.

L\'{e}vy walk dynamics describe enhanced transport phenomena in many systems. 
Within the CTRW framework, originally introduced by Montroll and Weiss \cite{MontrollWeiss:1965}, the significant feature of L\'{e}vy walk is the underlying spatiotemporal coupling, which penalizes long jumps and leads to a finite EAMSD \cite{ZaburdaevDenisovKlafter:2015}. While the uncoupled process, L\'{e}vy flight \cite{ShlesingerZaslavskyFrisch:1995,MetzlerKlafter:2000}, has divergent EAMSD.
The diffusion behavior of L\'{e}vy walk depends on the exponent $\alpha$ of the power-law distributed running time. It displays ballistic diffusion for $\alpha<1$ and sub-ballistic superdiffusion for $1<\alpha<2$.
We assume the particle switches between L\'{e}vy walk phase and Brownian phase, denoted as states `$+$' and `$-$', respectively.
The velocities of the two-state process are, respectively, $v_+(t)$ for L\'{e}vy walk and $v_-(t)$ for Brownian motion.
The PDF of $v_+(t)$ is $\delta(|v|-v_0)/2$, while $v_-(t)=\sqrt{2D}\xi(t)$ with $\xi(t)$ being a Gaussian white noise satisfying $\langle\xi(t)\rangle=0$ and $\langle\xi(t_1)\xi(t_2)\rangle=\delta(t_1-t_2)$. By taking the diffusivity $D=0$, the Brownian phase becomes a trap event and we immediately obtain the process -- L\'{e}vy walk interrupted by rest.

Let the sojourn times $t$ in the two states `$\pm$' be random variables obeying power-law distribution:
\begin{equation}
  \psi_\pm(t)\simeq \frac{a_\pm}{|\Gamma(-\alpha_\pm)|t^{1+\alpha_\pm}}
\end{equation}
for large $t$, where $a_\pm$ are scale factors and $\Gamma(\cdot)$ is the Gamma function. We assume that the exponents $\alpha_\pm\in(0,2)$ in two states and the sojourn times in two sates are mutually independent. As usual, we apply the approach of Laplace transform $\hat{\psi}_\pm(s):=\int_0^\infty dt e^{-st}\psi_\pm(t)$ and obtain the asymptotic behavior of the sojourn time distribution for small $s$:
\begin{align}
    &\hat{\psi}_\pm(s) \simeq 1-a_\pm s^{\alpha_\pm}   ~~\qquad\qquad {\rm for}~~ \alpha_\pm\in(0,1),  \\
    &\hat{\psi}_\pm(s) \simeq 1-\mu_\pm s + a_\pm s^{\alpha_\pm}   ~~~~~ {\rm for}~~ \alpha_\pm\in(1,2),
\end{align}
where $\mu_\pm$ is the mean sojourn time in state `$\pm$', being finite when $\alpha_\pm\in(1,2)$. The survival probability that the sojourn time in state `$\pm$' exceeds $t$ is defined as $\Psi_\pm(t)=\int_t^\infty dt' \psi_\pm(t')$ with Laplace transform $\hat{\Psi}_\pm(s)=[1-\hat{\psi}_\pm(s)]/s$. Note that the dynamic behaviors of standard L\'{e}vy walk are significantly different for exponent less or larger than $1$ \cite{ZaburdaevDenisovKlafter:2015}. We will fully discuss the EAMSD and TAMSD of the two-state process for different sets of $\alpha_\pm$ in the following. Although the mean sojourn time is finite (i.e., $\alpha_\pm>1$) in most cases, such as the intermittent search process, there are still some circumstances presenting scale free dynamics with $\alpha_\pm<1$, for example, the RNA transport in neuronal systems. Here we make uniform discussions with $\alpha_\pm\in(0,2)$ for comprehensive understanding of the two-state process.

\textit{Propagator and occupation fraction of two states}. Suppose that the particles are initialized at the origin. 
The propagator $p(x,t)$ represents the PDF of finding the particle at position $x$ at time $t$. For the two-state process, it is natural to concern which state the particles are located in at time $t$. Here we denote the joint PDF of finding the particle at position $x$ and state `$\pm$' at time $t$ as $p_\pm(x,t)$, which is associated with the propagator by the relation $p(x,t)=p_+(x, t)+p_-(x,t)$. The subscript `$\pm$' will imply an identical meaning for other quantities.

The integral equations for $p_\pm(x,t)$ can be similarly obtained as the master equations for CTRWs.
Besides the sojourn time distribution $\psi_\pm(t)$ and survival probability $\Psi_\pm(t)$, we introduce the notation $G_\pm(x,t)$ to represent the conditional probability that a particle makes a displacement $x$ during sojourn time $t$ at one step in state `$\pm$'. Their expressions are given by
\begin{align}
    &G_+(x,t)=\delta(|x|-v_0t)/2,  \\
    &G_-(x,t)=\frac{1}{\sqrt{4\pi Dt}}\exp\left(-\frac{x^2}{4Dt}\right),
\end{align}
since the state `$+$' represents L\'{e}vy walk and state `$-$' denotes Brownian motion, respectively. Then the transport equation governing flux of particles $\gamma_\pm(x,t)$, which defines how many particles leave the position $x$ and change from state `$\mp$' to state `$\pm$' per unit time, satisfies, 
\begin{equation}
\begin{split}
    \gamma_\pm(x,t)&= \int_0^tdt'\int_{-\infty}^{\infty}dx'\psi_\mp(t')G_\mp(x',t')\gamma_\mp(x-x',t-t')  \\
                 &~~~+p^0_\mp\psi_\mp(t)G_\mp(x,t),
\end{split}
\end{equation}
where the constant $p^0_\pm$ is the initial fraction of two states, that is $p_\pm(x,t=0)=p^0_\pm\delta(x)$.
The current density $p_\pm(x,t)$ of particles is connected to the flux $\gamma_\pm(x,t)$
\begin{equation}
  \begin{split}
  p_\pm(x,t)&=\int_0^t dt'\int_{-\infty}^\infty dx' \Psi_\pm(t')G_\pm(x',t')\gamma_\pm(x-x',t-t')  \\
            &~~~+p^0_\pm \Psi_\pm(t)G_\pm(x,t).
  \end{split}
\end{equation}

By means of the techniques of Laplace and Fourier transform, $\tilde{\hat{p}}_\pm(k,s)=\int_0^\infty dt\int_{-\infty}^\infty dx e^{-st}e^{ikx}p_\pm(x,t)$ can be obtained (see Supplemental Material). Besides, the occupation fraction of two states $p_\pm(t)$, as the marginal density of finding the particles in state `$\pm$' at time $t$, can be obtained by taking $k = 0$ in $\tilde{\hat{p}}_\pm(k,s)$. The expression of $p_\pm(t)$ in Laplace space $(t\rightarrow s)$ is
\begin{equation}\label{ps}
  \hat{p}_{\pm}(s)=\frac{p_{\pm}^0+p_{\mp}^0\hat{\psi}_{\mp}(s)}{1-\hat{\psi}_+(s)\hat{\psi}_-(s)}
  \cdot\frac{1-\hat{\psi}_{\pm}(s)}{s},
\end{equation}
the normalization of which can be confirmed by verifying $\hat{p}_+(s)+\hat{p}_-(s)=1/s$.

\textit{EAMSD and TAMSD}. If one is eager for more information of a process, such as the TAMSD, the propagator $p(x,t)$ at a single point is not enough. Instead, the two-point velocity correlation function $\langle v(t_1)v(t_2)\rangle$ plays a crucial role. We will calculate it firstly and then show the generic results of EAMSD and TAMSD for the aging process $x_{t_a}(t)$. The age $t_a$ means that this process has evolved for a time period $t_a$ before we start to observe it, and $t$ is the measurement time.

Since the model we considered contains two states: L\'{e}vy walk and Brownian motion, represented by symbols `$+$' and `$-$', respectively. The velocity correlation function could be written as a sum of four possible cases in terms of different states:
\begin{equation}\label{vv}
\begin{split}
\langle v(t_1)v(t_2) \rangle&=\langle v_+(t_1)v_+(t_2) \rangle+\langle v_-(t_1)v_-(t_2) \rangle\\
                            &~~~+\langle v_+(t_1)v_-(t_2) \rangle+\langle v_-(t_1)v_+(t_2) \rangle.
\end{split}
\end{equation}
The first term on the right-hand side represents the case that the velocity process $v(t)$ are in L\'{e}vy walk phase at both time points $t_1$ and $t_2$; other terms stand for similar parts of the correlation function. 
For the first term, the velocity is correlated only when there is no renewal happens between $t_1$ and $t_2$. Thus, we have
\begin{equation}\label{vv1}
  \langle v_+(t_1)v_+(t_2) \rangle = v_0^2 p_+(t_1) p_{+,0}(t_1,t_2),
\end{equation}
where $p_+(t)$ has been given in \eqref{ps} and $p_{+,0}(t_1,t_2)$ is the PDF that no renewal happens between times $t_1$ and $t_2$ in state `$+$'. Similarly, the second term on the right hand side of \eqref{vv} is
\begin{equation}\label{vv2}
\begin{split}
    \langle v_-(t_1)v_-(t_2) \rangle = 2D\delta(t_1-t_2) p_-(t_1) p_{-,0}(t_1,t_2),
\end{split}
\end{equation}
where $p_{-,0}(t_1,t_2)=1$ for $t_1=t_2$, since there must be no renewals within a zero time lag. The two states at times $t_1$ and $t_2$ are different in the last two terms on \eqref{vv}. Therefore, the velocity at $t_1$ and $t_2$ are independent. Considering the velocity is unbiased at any time, the last two terms are void.

Note that the PDFs $p_\pm(t)$ and $p_{+,0}(t_1,t_2)$ should be calculated firstly to obtain the velocity correlation function in \eqref{vv}. The former one has been given in \eqref{ps}, while the double Laplace transform $(t\rightarrow s,\tau\rightarrow u)$ of the latter PDF $f_{+,0}(t,\tau)=p_{+,0}(t,t+\tau)$ is \cite{GodrecheLuck:2001}
\begin{equation}\label{ps1s20}
\hat{f}_{+,0}(s,u)= \frac{s-u+u\hat{\psi}_{+}(s)-s\hat{\psi}_{+}(u)}{s(s-u)(1-\hat{\psi}_{+}(s))u}.
\end{equation}
It seems not easy to perform the inverse Laplace transform on $\hat{f}_{+,0}(s,u)$. Instead, we can obtain the expression of $p_{+,0}(t_1, t_2)$ in Laplace space ($t_1\rightarrow s_1, t_2\rightarrow s_2$) by substituting variables  (see Supplemental Material):
\begin{equation}\label{ps1s2}
\begin{split}
\hat{p}_{+,0}(s_1,s_2)=  \frac{1+\hat{\psi}_+(s_1+s_2)-\hat{\psi}_+(s_1)-\hat{\psi}_+(s_2)}{s_1s_2(1-\hat{\psi}_+(s_1+s_2))}.
\end{split}
\end{equation}
Taking inverse Laplace transform on \eqref{ps1s2} becomes doable.
Based on \eqref{ps} and \eqref{ps1s2}, the velocity correlation function $\langle v(t)v(t+\tau)\rangle$ in \eqref{vv} can be obtained for different sojourn time distributions $\psi_\pm(t)$. Noticing the asymptotic forms of $p_\pm(t)$ and $p_{+,0}(t,t+\tau)$ for large $t$, the velocity correlation function can be rewritten in the scaling form as
\begin{equation}\label{vtvttau}
\begin{split}
\langle v(t)v(t+\tau)\rangle &=\langle v_+(t)v_+(t+\tau)\rangle+\langle v_-(t)v_-(t+\tau)\rangle\\
                            &\simeq C_1t^{\nu_1-2}\rho\left(\frac{\tau}{t}\right)+C_2t^{\nu_2-1}\delta(\tau),
\end{split}
\end{equation}
where the parameters $\nu_1,\nu_2$ and the scaling function $\rho(\cdot)$ are determined by $p_\pm(t)$ and $p_{+,0}(t,t+\tau)$. The scaling form \eqref{vtvttau} helps to show different scaling behaviors of $\langle v(t)v(t+\tau)\rangle$ for different sojourn time distributions $\psi_\pm(t)$, and brings convenience to give a generic expressions of MSDs \cite{DechantLutzKesslerBarkai:2014,MeyerBarkaiKantz:2017}.

Now we focus on the aging process $x_{t_a}(t)$. The EAMSD of this aging process is defined as $\langle x_{t_a}^2(t)\rangle=\langle (x(t_a+t)-x(t_a))^2\rangle$, which can be obtained through the scaling form in \eqref{vtvttau}. For weak aging $t_a\ll t$ and strong aging $t_a\gg t$ cases  (see Supplemental Material), it behaves as
\begin{equation}\label{EA}
  \langle x_{t_a}^2(t)\rangle\simeq
  \left\{
    \begin{array}{ll}
      K_1 t^{\nu_1}+K_2 t^{\nu_2}, & t_a\ll t, \\
      K_3 t_a^\beta t^{\nu_1-\beta}+C_2 t_a^{\nu_2-1}t, & t_a\gg t,
    \end{array}
  \right.
\end{equation}
where the coefficients $K_1=2C_1/\nu_1\int_0^\infty dt (t+1)^{-\nu_1 }\rho(t)$, $K_2=C_2/\nu_2$ and $K_3=2c_1C_1[(\nu_1-\beta-1)(\nu_1-\beta)]^{-1}$.
Here $c_1$ depends on the asymptotic form of scaling function $\rho(z)\simeq c_1 z^{-\delta_1}$ for small $z$, and
$\beta$ is the exponent of the variance of velocity in the L\'{e}vy walk phase for large $t$ \cite{DechantLutzKesslerBarkai:2014}, i.e.,
\begin{equation}
  \langle v_+^2(t)\rangle=v_0^2p_+(t)\propto t^{\beta}.
\end{equation}

When constructing single particle tracking experiments, the aging process $x_{t_a}(t)$ is evaluated in terms of its TAMSD, which is defined as $\overline{\delta_{t_a}^2(\Delta)}=1/(T-\Delta)\int_{t_a}^{t_a+T-\Delta}dt [x(t+\Delta)-x(t)]^2$ with $\Delta$ denoting the lag time and $T$ the total measurement time \cite{MetzlerJeonCherstvyBarkai:2014}. The TAMSD is calculated in the limit $\Delta\ll T$ to obtain good statistics. 
Weak ergodicity breaking is the common phenomenon of a majority of anomalous diffusion. Similarly to the procedure of calculating EAMSD, we obtain the ensemble-averaged TAMSD as (see Supplemental Material):
\begin{equation}\label{TA}
  \langle\overline{\delta_{t_a}^2(\Delta)}\rangle\simeq
  \left\{
    \begin{array}{ll}
      \frac{K_3}{1+\beta} \,T^{\beta}\Delta^{\nu_1-\beta}+K_2 T^{\nu_2-1}\Delta, & t_a\ll T, \\
      K_3 t_a^\beta \Delta^{\nu_1-\beta}+C_2 t_a^{\nu_2-1}\Delta, & t_a\gg T.
    \end{array}
  \right.
\end{equation}

There are at least four findings being worth to report from the observations of the generic results of EAMSDs in \eqref{EA} and TAMSDs in \eqref{TA}.
(i) All the four mentioned formulae consist of two parts (one from L\'{e}vy walk phase and another one from Brownian phase). The exponents of evolution time $t$ or time lag $\Delta$ in these two parts might be different from the ones of the corresponding individual L\'{e}vy walk and Brownian motion. This is because the PDF $p_\pm(t)$ in \eqref{ps} plays a weighted role on L\'{e}vy walk and Brownian motion. Besides, the sums of exponents of the time variables (including $t,t_a,T,\Delta$) in individual two parts are $\nu_1$ and $\nu_2$, respectively, whatever it is EAMSD or TAMSD, and weak or strong aging cases.
(ii) The exponents of time variables in weak and strong aging cases are closely related for TAMSD in \eqref{TA}. While keeping the exponents of $\Delta$ invariant and replacing measurement time $T$ by age $t_a$, the result for strong aging case is obtained from the one of weak aging case.
In other words, the TAMSD for weak aging case only depends on $T$ and $\Delta$, while in the same way it counts on $t_a$ and $\Delta$ for strong aging cases.
(iii) The EAMSD and TAMSD in weak aging case do not depend on the age $t_a$, the results of which are identical to the non-aging case $t_a = 0$. In contrast, they explicitly depend on $t_a$ for strong aging case, which implies that the exponents $\beta$ and $\nu_2-1$ of $t_a$ must be zero if the equilibrium initial ensemble (i.e., $t_a\rightarrow\infty$ discussed in last section) of this system exists (see specific case $2$ in Table \ref{table}). And in this case, the TAMSD will be the same for weak and strong aging cases, and only depends on $\Delta$.
(iiii) Comparing the strong aging EAMSD and the mean of TAMSD \eqref{TA}, it can be noted that
\begin{equation}
  \langle x_{t_a}^2(\Delta)\rangle = \langle\overline{\delta_{t_a}^2(\Delta)}\rangle \quad {\rm for }~ t_a\gg T,
\end{equation}
which shows that the aging seemingly makes the weak ergodicity breaking system to be ergodic. It is clear that for any $\alpha_-$ Brownian motion is ergodic in its own phase. However, for TAMSD in L\'{e}vy walk phase, there are some differences between $\alpha_+<1$ and $1<\alpha_+<2$. For $1<\alpha_+<2$, the mean sojourn time in L\'{e}vy walk phase is finite, individual trajectories become self-averaging at sufficiently long (infinite) times, such that there will be no difference between $\overline{\delta_{t_a}^2(\Delta)}$ obtained from different trajectories and ensemble-averaged quantity $\langle\overline{\delta_{t_a}^2(\Delta)}\rangle$ \cite{GodecMetzler:2013,FroembergBarkai:2013}. While for $\alpha_+<1$, the characteristic time scale is infinite, then the individual TAMSD $\overline{\delta_{t_a}^2(\Delta)}$ is irreproducible and inequivalent with the corresponding EAMSD.

\textit{Specific cases}. Since both $\alpha_+$ and $\alpha_-$ go through the range $(0,2)$, it can be divided into six cases as shown in Table \ref{table}. See the detailed derivations of parameters $\nu_1,\nu_2$, and $\beta$ for these cases in (Supplemental Material).
It seems tedious to discuss the EAMSDs and TAMSDs individually for six different cases of $\alpha_\pm$. In fact, they can be organized into three categories to deepen understandings of the two-state process by considering the properties of its ingredients --- L\'{e}vy walk and Brownian motion. It is well-known that the standard L\'{e}vy walk performs ballistic diffusion when the exponent of the distribution of running times $\alpha<1$ and sub-ballistic superdiffusion when $1<\alpha<2$, which is faster than the normal diffusion of Brownian motion. Based on this understanding, the Brownian phase undoubtedly suppresses the diffusion behavior of L\'{e}vy walk. This effect may be durable or transient, which is completely determined by the fraction of two states $p_\pm(t)$, or more essentially, the magnitude of the exponents $\alpha_\pm$. From this point of view, the three categories are: (i) $\alpha_+$ and $\alpha_-$ are comparable, including the first two cases in Table \ref{table}; (ii) $\alpha_+$ is smaller, including the middle two cases in Table \ref{table}; (iii) $\alpha_-$ is smaller, including the last two cases in Table \ref{table}.

\begin{table}
\centering
\caption{Values of several major parameters of EAMSD and TAMSD in \eqref{EA} and \eqref{TA} for six cases with different $\alpha_\pm$.}\label{table}
\begin{tabular}{|l|l|l|l|}
  \hline
  specific cases & $\nu_1$ & $\nu_2$ & $\beta$ \\
\hline
  1. $\alpha_+=\alpha_-<1$ & $2$ & $1$ & $0$ \\
  2. $1<\alpha_\pm<2$ & $3-\alpha_+$ & $1$ & $0$ \\
  3. $\alpha_+<\alpha_-<1$ & $2$ & $\alpha_+-\alpha_-+1$ & $0$ \\
  4. $\alpha_+ < 1 < \alpha_- < 2$ & $2$ & $\alpha_+$ & $0$ \\
  5. $\alpha_-<\alpha_+<1$ & $\alpha_--\alpha_++2$ & $1$ & $\alpha_--\alpha_+$ \\
  6. $\alpha_-<1<\alpha_+<2$ & $\alpha_--\alpha_++2$ & $1$ & $\alpha_--1$ \\
  \hline
\end{tabular}
\end{table}

\begin{figure*}[tbhp]
\begin{minipage}{0.31\linewidth}
\includegraphics[scale=0.36]{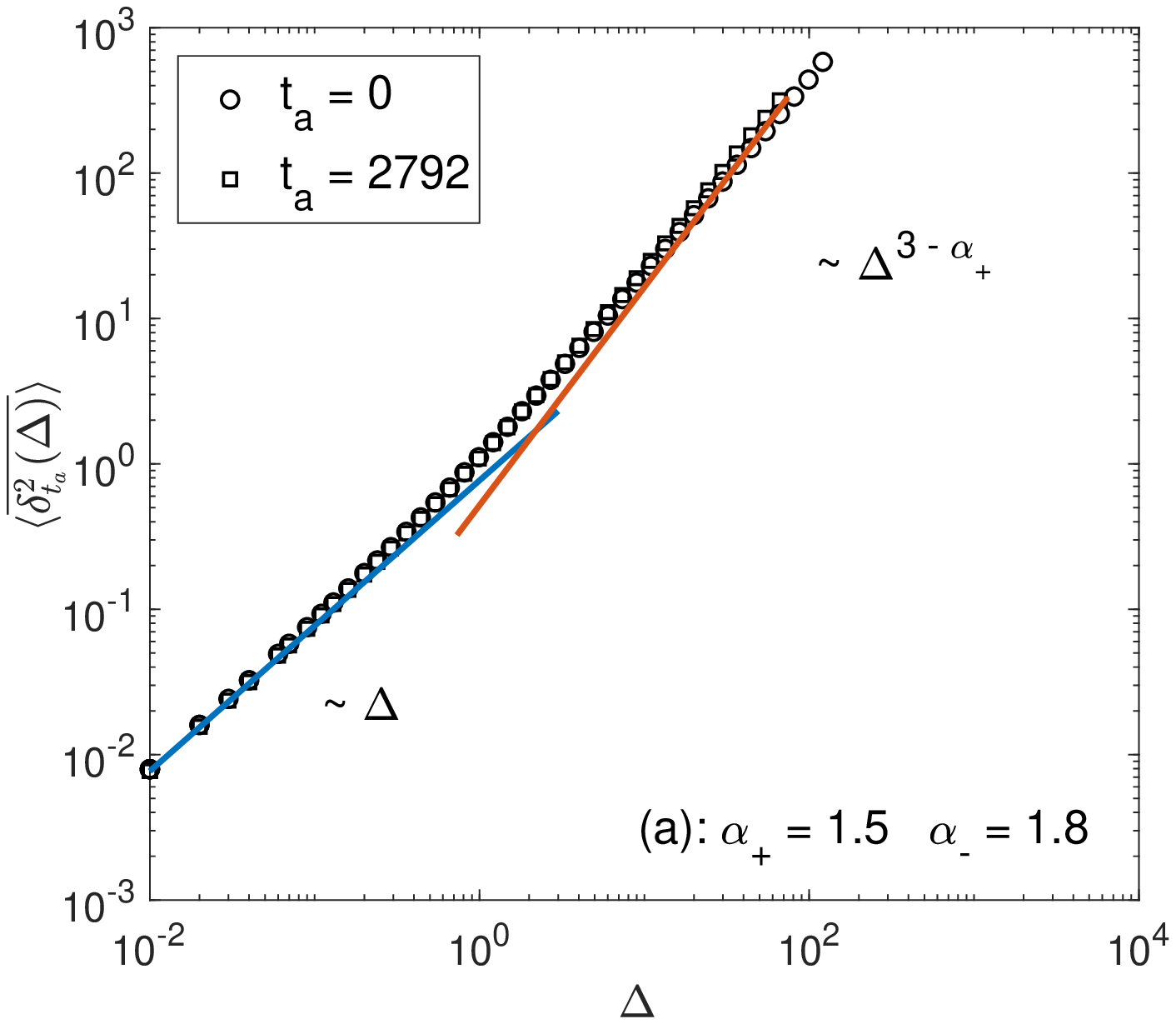}
\end{minipage}
\begin{minipage}{0.31\linewidth}
\includegraphics[scale=0.36]{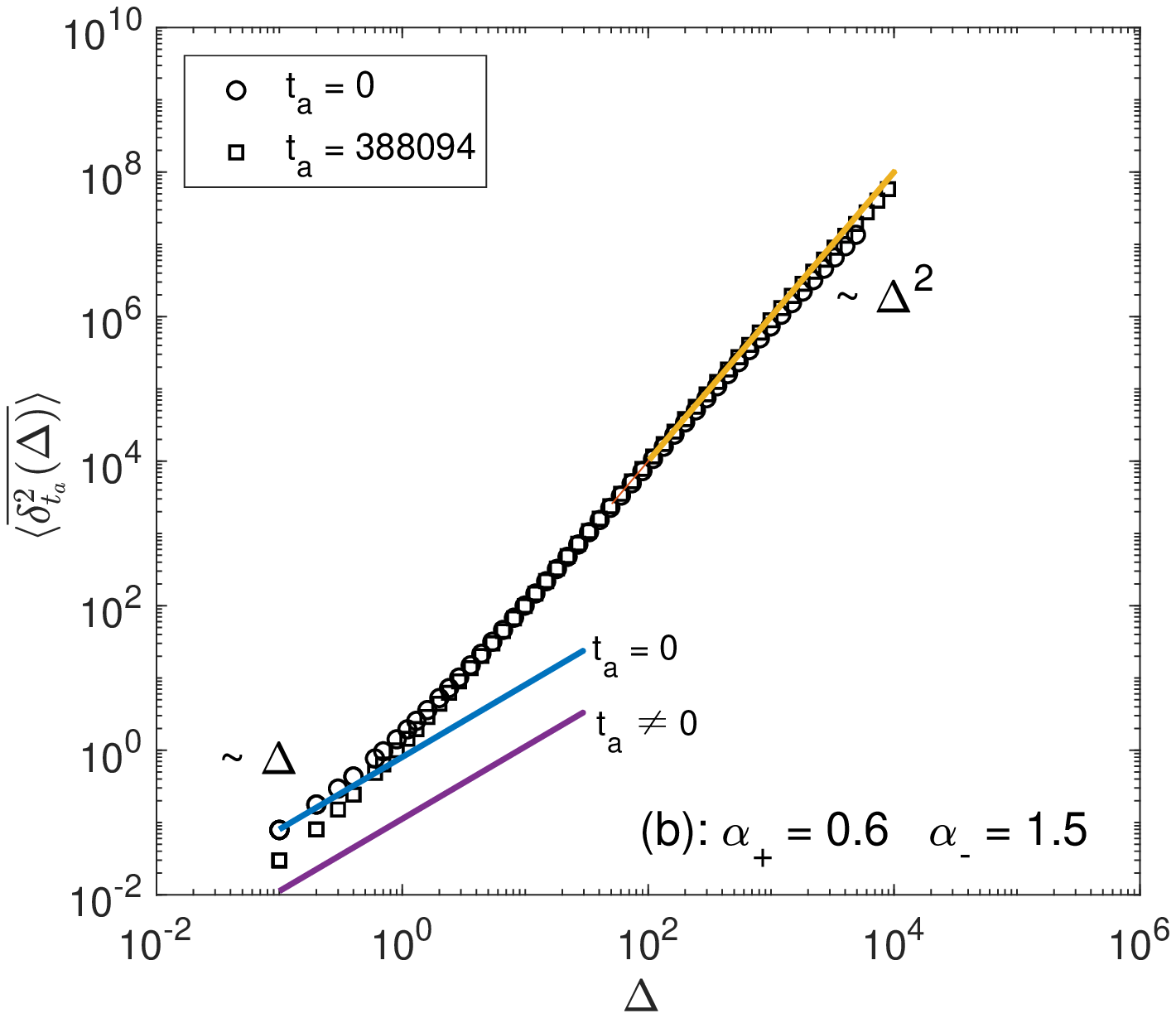}
\end{minipage}
\begin{minipage}{0.31\linewidth}
\includegraphics[scale=0.36]{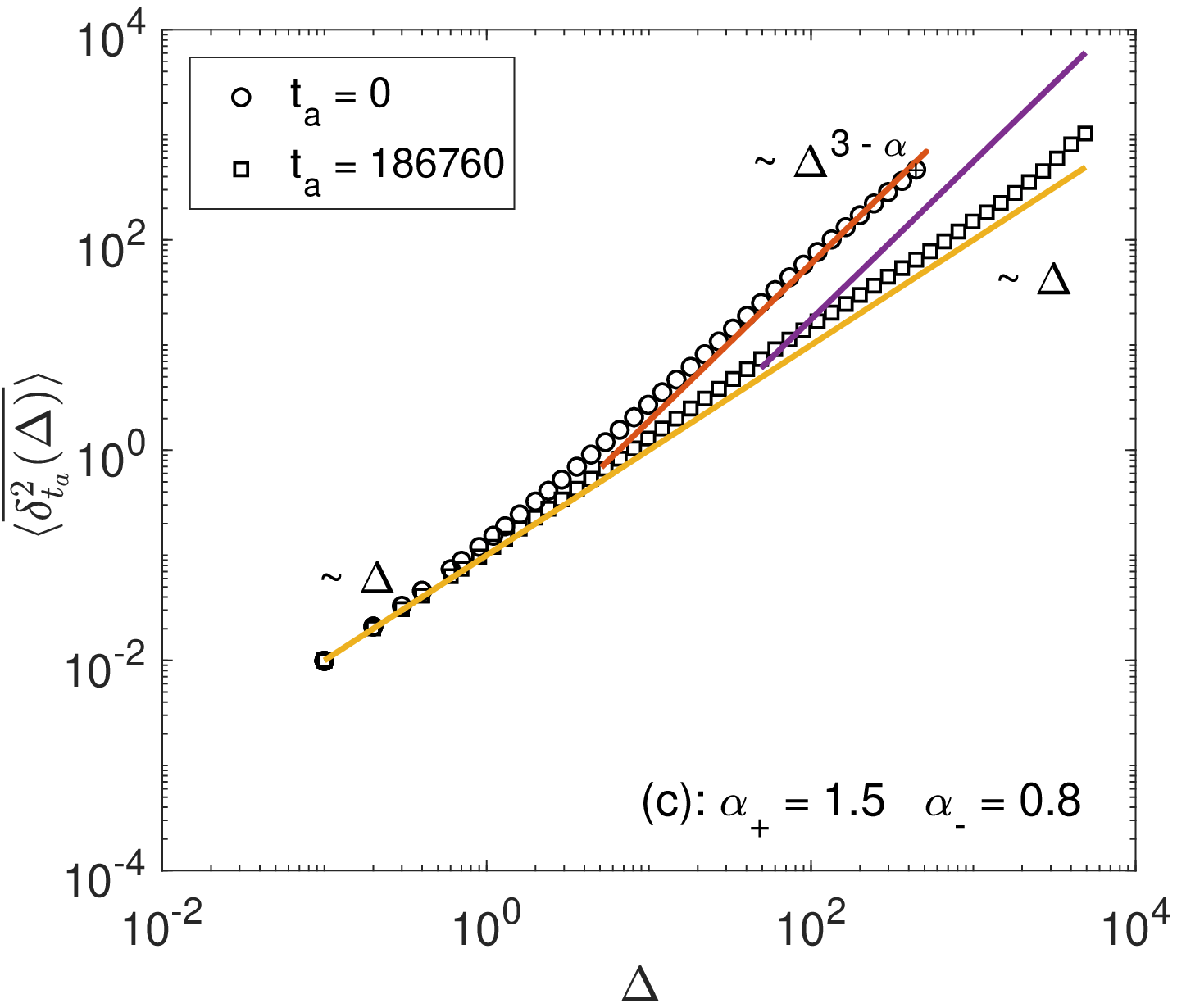}
\end{minipage}
\caption{TAMSD of the two-state process for different sets of $\alpha_\pm$. Black circles and squares represent the simulation results of the mean value of TAMSD averaging over 200 realizations, and the solid lines are the theoretical ones (with small and large asymptotic forms in \eqref{TA}). (a): case (i) with $\alpha_+=1.5,\alpha_-=1.8$ and measurement time $T=200$. The simulation results agree with the theoretical ones for small time $(\sim\Delta)$ and large time $(\sim\Delta^{3-\alpha_+})$. (b): case (ii) with $\alpha_+=0.6,\alpha_-=1.5$ and measurement time $T=20426$. The asymptotic behavior $\sim\Delta^2$ for large time is observed. The simulation and theoretical results do not coincide for small $\Delta$ in strong aging case, since the coefficient $t_a^{\alpha_+-1}$ in front of $\Delta$ in \eqref{case2} is too small and another term $\Delta^2$ dominates. (c): case (iii) with $\alpha_+=1.5,\alpha_-=0.8$ and measurement time $T=9830$. The asymptotic behavior $\sim\Delta$ for short time can be observed. It does not coincide for large $\Delta$ in strong aging case, since the coefficient $t_a^{\alpha_--1}$ in front of $\Delta^{3-\alpha_+}$ in \eqref{case3} is too small and another term $\Delta$ dominates. Therefore, there is not much difference between the strong aging simulations and the whole solid (yellow) line with slope $1$ for large $\Delta$.}\label{figure}
\end{figure*}

As representatives of the above three situations, we choose three sets of parameters: (i) $\alpha_+=1.5,\alpha_-=1.8$, (ii) $\alpha_+=0.6,\alpha_-=1.5$, and (iii) $\alpha_+=1.5,\alpha_-=0.8$. The corresponding TAMSDs for weak and strong aging cases are simulated and shown in Fig. \ref{figure}. The TAMSDs for other cases and EAMSDs are presented in (Supplemental Material). The theoretical TAMSDs for these three cases can be obtained from \eqref{TA} as:
\begin{equation}\label{case1}
 \textrm{(i)}~ \langle \overline{\delta^2_{t_a}(\Delta)}\rangle \simeq
  \left\{
    \begin{array}{ll}
      2D_2\Delta^{3-\alpha_+}+\frac{2D\mu_-}{\mu_++\mu_-}\Delta, & t_a\ll T, \\
      2D_2\Delta^{3-\alpha_+}+\frac{2D\mu_-}{\mu_++\mu_-}\Delta, & T\ll t_a,
    \end{array}
  \right.
\end{equation}
\begin{equation}\label{case2}
  \textrm{(ii)}~\langle \overline{\delta^2_{t_a}(\Delta)}\rangle \simeq
  \left\{
    \begin{array}{ll}
      v_0^2\Delta^2+\frac{2D\mu_-}{a_+\Gamma(1+\alpha_+)}T^{\alpha_+-1}\Delta, & t_a\ll T, \\
      v_0^2\Delta^2+\frac{2D\mu_-}{a_+\Gamma(\alpha_+)}t_a^{\alpha_+-1}\Delta, & T\ll t_a,
    \end{array}
  \right.
\end{equation}
\begin{equation}\label{case3}
  \textrm{(iii)}~\langle \overline{\delta^2_{t_a}(\Delta)}\rangle \simeq
  \left\{
    \begin{array}{ll}
      2D_4T^{\alpha_--1}\Delta^{3-\alpha_+}+2D\Delta, & t_a\ll T, \\
      2D_4\alpha_-t_a^{\alpha_--1}\Delta^{3-\alpha_+}+2D\Delta, & T\ll t_a.
    \end{array}
  \right.
\end{equation}
For the first category (i), a stationary of the fractions of two states $p_\pm(t)$ can be achieved for long times, that is, $p_\pm(t)$ tends to a constant not equal to $0$ or $1$ (see Supplemental Material). Then the EAMSD and TAMSD are the combination of the fraction of analogues of individual L\'{e}vy walk and Brownian motion whether it is weak aging or strong aging.
For the second category (ii) with $\alpha_+<\alpha_-$ where $p_+(t)\rightarrow1$ as $t\rightarrow\infty$, the L\'{e}vy walk phase in state `$+$' tends to occupy the whole time. Then the results are naturally similar to an individual L\'{e}vy walk, except for the small asymptotic form $\Delta$ resulting from Brownian phase.
For the third category (iii) with $\alpha_+>\alpha_-$, by contrast, now $p_-(t)\rightarrow1$ as $t\rightarrow\infty$ and the L\'{e}vy walk phase in state `$+$' gradually withdraws from the two states in a power-law way. This power-law way suppresses the diffusion of L\'{e}vy walk phase and gives the opportunity to Brownian motion to be the leading term when $\alpha_+-\alpha_->1$.
In conclusion, compared to the EAMSD and TAMSD of individual aging L\'{e}vy walk \cite{MagdziarzZorawik:2017} and Brownian motion, it can be found that the fraction $p_\pm(t)$ in a two-state process plays a crucial role. It contributes a power term of $\Delta$ to weak aging EAMSD, a power term of $T$ to weak aging TAMSD, and a power term of $t_a$ to strong aging EAMSD and TAMSD.

The model L\'{e}vy walk interrupted by rest has attracted considerable attention in physics \cite{SolomonWeeksSwinney:1993,KlafterZumofen:1994} and biology \cite{SongMoonJeonPark:2018}. 
The EAMSD and TAMSD for this model can be obtained by taking the diffusivity $D$ in Brownian phase to be zero. It has been pointed that all the results above consist two parts corresponding to L\'{e}vy walk and Brownian motion, respectively. Taking $D = 0$ just eliminates the latter part and brings no effect on the former part of L\'{e}vy walk phase. For L\'{e}vy walk interrupted by rest, the asymptotic behavior of small $\Delta$ in TAMSD disappears and subdiffusion behavior might exist if $\alpha_+-\alpha_->1$.

\textit{Initial ensemble}. In general, the standard L\'{e}vy walk model is a non-Markovian process and so is the two-state process alternating between L\'{e}vy walk and Brownian motion with power-law distributed sojourn time.
It is natural to consider the effects of the initial ensembles of the particles. 
It is called a nonequilibrium initial ensemble \cite{Cox:1962,KlafterZumofen:1993} if all particles are introduced to the system at $t = 0$ without any prehistories. In contrast, if the particles have been evolving for time $t_1$ before we start to measure this system, we call this system with equilibrium initial ensemble when $t_1\rightarrow\infty$ \cite{Cox:1962,KlafterZumofen:1993}. The EAMSD of standard L\'{e}vy walk has been shown to be different for different initial ensemble \cite{ZaburdaevDenisovKlafter:2015,WangChenDeng:2019}. Note that the equilibrium initial ensemble exists only if the sojourn times in two states `$\pm$' both have finite first moments, i.e., $1 <\alpha_\pm < 2$ in our concerned model.

For nonequilibrium initial ensemble, the corresponding EAMSD $\langle x^2(\Delta)\rangle$ and TAMSD $\overline{\delta^2(\Delta)}$ can be obtained by taking $t_a=0$ in previous section, i.e.,
\begin{equation}
  \begin{split}
    \langle x^2(\Delta)\rangle = \langle x_{t_a}^2(\Delta)\rangle |_{t_a=0},  \quad
    \overline{\delta^2(\Delta)} = \overline{\delta_{t_a}^2(\Delta)} |_{t_a=0}.
  \end{split}
\end{equation}
Since the results of the weak aging case (i.e., $t_a\ll\Delta$) with different sojourn time pairs $\psi_\pm(t)$ in Eqs. (\ref{EA}) and (\ref{TA}) are independent of $t_a$, they are indeed the results for nonequilibrium initial ensemble. When $1<\alpha_\pm<2$, the results of the strong aging case (i.e., $t_a\gg T$) in \eqref{case1} are independent of $t_a$. Therefore, the EAMSD $\langle x^2_{\textrm{eq}}(\Delta)\rangle$ and TAMSD $\langle\overline{\delta^2_{\textrm{eq}}(\Delta)}\rangle$ for equilibrium initial ensemble ($t_a\rightarrow\infty$) are
\begin{equation}
  \langle x^2_{\textrm{eq}}(\Delta)\rangle=\langle\overline{\delta^2_{\textrm{eq}}(\Delta)}\rangle
  \simeq2D_2\Delta^{3-\alpha_+}+ \frac{2D\mu_-}{\mu_++\mu_-}\Delta.
\end{equation}
If the sojourn times are so long that the mean sojourn time diverges, there is no sense in talking about the equilibrated initial ensemble. However, the asymptotic behaviors of strong aging case $t_a\gg\Delta$ can still be investigated (see Supplemental Material). There is a special case $0<\alpha_+=\alpha_-<1$, where the particles reach a balance that each half of them are located in each of the two states and the EAMSD and TAMSD are both independent on the age $t_a$, that is,
\begin{equation}
  x^2_{t_a}(\Delta)\rangle = \langle\overline{\delta^2_{t_a}(\Delta)}\rangle
  \simeq \frac{v_0^2}{2}\Delta^2 +D\Delta
\end{equation}
for sufficiently large $t_a$. If $\alpha_+\neq\alpha_-$ and at least one of them less than $1$, then neither an equilibrium initial ensemble nor a balance for long time exists. The state with small exponent $\alpha_\pm$ of sojourn time distribution will dominate the MSD for long times. One can see this phenomenon in the last four cases in Table \ref{table}. In these cases, the EAMSD and TAMSD for strong aging cases all consist of two parts corresponding to L\'{e}vy walk and Brownian motion. One of the parts is independent on $t_a$ while another part contains a power term of $t_a$ with a negative exponent. The latter part tends to zero as $t_a\rightarrow\infty$ and the former one dominates, which corresponds to the state with smaller exponent $\alpha_\pm$ of sojourn time distributions.

\textit{Conclusion}. It often happens that a single-state process cannot sufficiently describe the observed physical and biological phenomena. Two-state process is a kind of simple but important model to characterize some of these phenomena. A Langevin equation with two diffusion modes (fast and slow diffusion modes) has been investigated in \cite{MiyaguchiAkimotoYamamoto:2016}, where a transient subdiffusion and the non-Gaussian propagator for short time are observed for a nonequilibrium ensemble.
In this report, we consider a two-state process with fast phase (L\'{e}vy walk) and slow phase (Brownian motion), which is also the intermittent search process for finding rare hidden targets. It is not easy to model the process with two completely different phases by a Langevin equation.
By contrast, we resort to the velocity process $v(t)$, which also consists of two states. Based on the velocity correlation function, we obtain the generic expressions of the EAMSD and TAMSD for different sojourn time distributions.

One of the key contributions of this report is to explicitly discuss the relation between EAMSD and TAMSD.
In particular, the weak and strong aging cases are also considered for these MSDs since the measurement in experiments might not begin at the start of the process concerned. It is found that the occupation fraction plays a weighed role in L\'{e}vy walk phase and Brownian phase, and the MSDs are just a combination of these two parts. 
The meticulous discussions on the aging MSDs are helpful to understand the two-state process and to analyze the experimental data.

If taking the diffusivity $D$ to be zero in Brownian phase, we obtain another important process --- L\'{e}vy walk interrupted by rest. 
Taking $D=0$ just eliminates the contributions from Brownian phase.
From another aspect of the two-state process, we find the fact that the slow phase, whether it is rest or Brownian motion, suppresses the diffusion behavior of L\'{e}vy walk if its sojourn time is longer than that of L\'{e}vy walk phase. The mechanism is similar to the trap event \cite{GoldingCox:2006} in CTRW models.
Compared to them, there exist some other models describing the suppression of  the diffusion of L\'{e}vy walk with different mechanism, such as the L\'{e}vy walk with memory in running time \cite{ChenWangDeng:2019} and the walker moving in a heterogeneous medium \cite{KaminskaSrokowski:2018}.

\textit{Acknowledgments}. This work was supported by the National Natural Science Foundation of China under grant no. 11671182, and the Fundamental Research Funds for the Central Universities under grant no. lzujbky-2018-ot03.

\bibliography{Reference}

\begin{thebibliography}{34}%
\makeatletter
\providecommand \@ifxundefined [1]{%
 \@ifx{#1\undefined}
}%
\providecommand \@ifnum [1]{%
 \ifnum #1\expandafter \@firstoftwo
 \else \expandafter \@secondoftwo
 \fi
}%
\providecommand \@ifx [1]{%
 \ifx #1\expandafter \@firstoftwo
 \else \expandafter \@secondoftwo
 \fi
}%
\providecommand \natexlab [1]{#1}%
\providecommand \enquote  [1]{``#1''}%
\providecommand \bibnamefont  [1]{#1}%
\providecommand \bibfnamefont [1]{#1}%
\providecommand \citenamefont [1]{#1}%
\providecommand \href@noop [0]{\@secondoftwo}%
\providecommand \href [0]{\begingroup \@sanitize@url \@href}%
\providecommand \@href[1]{\@@startlink{#1}\@@href}%
\providecommand \@@href[1]{\endgroup#1\@@endlink}%
\providecommand \@sanitize@url [0]{\catcode `\\12\catcode `\$12\catcode
  `\&12\catcode `\#12\catcode `\^12\catcode `\_12\catcode `\%12\relax}%
\providecommand \@@startlink[1]{}%
\providecommand \@@endlink[0]{}%
\providecommand \url  [0]{\begingroup\@sanitize@url \@url }%
\providecommand \@url [1]{\endgroup\@href {#1}{\urlprefix }}%
\providecommand \urlprefix  [0]{URL }%
\providecommand \Eprint [0]{\href }%
\providecommand \doibase [0]{https://doi.org/}%
\providecommand \selectlanguage [0]{\@gobble}%
\providecommand \bibinfo  [0]{\@secondoftwo}%
\providecommand \bibfield  [0]{\@secondoftwo}%
\providecommand \translation [1]{[#1]}%
\providecommand \BibitemOpen [0]{}%
\providecommand \bibitemStop [0]{}%
\providecommand \bibitemNoStop [0]{.\EOS\space}%
\providecommand \EOS [0]{\spacefactor3000\relax}%
\providecommand \BibitemShut  [1]{\csname bibitem#1\endcsname}%
\let\auto@bib@innerbib\@empty
\bibitem [{\citenamefont {Bell}(1991)}]{Bell:1991}%
  \BibitemOpen
  \bibfield  {author} {\bibinfo {author} {\bibfnamefont {J.~W.}\ \bibnamefont
  {Bell}},\ }\href@noop {} {\emph {\bibinfo {title} {Searching Behaviour, the
  Behavioural Ecology of Finding Resources}}}\ (\bibinfo  {publisher} {Chapman
  and Hall},\ \bibinfo {address} {London},\ \bibinfo {year} {1991})\BibitemShut
  {NoStop}%
\bibitem [{\citenamefont {B\'{e}nichou}\ \emph {et~al.}(2011)\citenamefont
  {B\'{e}nichou}, \citenamefont {Loverdo}, \citenamefont {Moreau},\ and\
  \citenamefont {Voituriez}}]{BenichouLoverdoMoreauVoituriez:2011}%
  \BibitemOpen
  \bibfield  {author} {\bibinfo {author} {\bibfnamefont {O.}~\bibnamefont
  {B\'{e}nichou}}, \bibinfo {author} {\bibfnamefont {C.}~\bibnamefont
  {Loverdo}}, \bibinfo {author} {\bibfnamefont {M.}~\bibnamefont {Moreau}},\
  and\ \bibinfo {author} {\bibfnamefont {R.}~\bibnamefont {Voituriez}},\
  }\bibfield  {title} {\bibinfo {title} {Intermittent search strategies},\
  }\href@noop {} {\bibfield  {journal} {\bibinfo  {journal} {Rev. Mod. Phys.}\
  }\textbf {\bibinfo {volume} {83}},\ \bibinfo {pages} {81} (\bibinfo {year}
  {2011})}\BibitemShut {NoStop}%
\bibitem [{\citenamefont {B\'{e}nichou}\ \emph {et~al.}(2005)\citenamefont
  {B\'{e}nichou}, \citenamefont {Coppey}, \citenamefont {Moreau}, \citenamefont
  {Suet},\ and\ \citenamefont
  {Voituriez}}]{BenichouCoppeyMoreauSuetVoituriez:2005}%
  \BibitemOpen
  \bibfield  {author} {\bibinfo {author} {\bibfnamefont {O.}~\bibnamefont
  {B\'{e}nichou}}, \bibinfo {author} {\bibfnamefont {M.}~\bibnamefont
  {Coppey}}, \bibinfo {author} {\bibfnamefont {M.}~\bibnamefont {Moreau}},
  \bibinfo {author} {\bibfnamefont {P.-H.}\ \bibnamefont {Suet}},\ and\
  \bibinfo {author} {\bibfnamefont {R.}~\bibnamefont {Voituriez}},\ }\bibfield
  {title} {\bibinfo {title} {Optimal search strategies for hidden targets},\
  }\href@noop {} {\bibfield  {journal} {\bibinfo  {journal} {Phys. Rev. Lett.}\
  }\textbf {\bibinfo {volume} {94}},\ \bibinfo {pages} {198101} (\bibinfo
  {year} {2005})}\BibitemShut {NoStop}%
\bibitem [{\citenamefont {Lomholt}\ \emph {et~al.}(2008)\citenamefont
  {Lomholt}, \citenamefont {Koren}, \citenamefont {Metzler},\ and\
  \citenamefont {Klafter}}]{LomholtKorenMetzlerKlafter:2008}%
  \BibitemOpen
  \bibfield  {author} {\bibinfo {author} {\bibfnamefont {M.~A.}\ \bibnamefont
  {Lomholt}}, \bibinfo {author} {\bibfnamefont {T.}~\bibnamefont {Koren}},
  \bibinfo {author} {\bibfnamefont {R.}~\bibnamefont {Metzler}},\ and\ \bibinfo
  {author} {\bibfnamefont {J.}~\bibnamefont {Klafter}},\ }\bibfield  {title}
  {\bibinfo {title} {L\'{e}vy strategies in intermittent search processes are
  advantageous},\ }\href@noop {} {\bibfield  {journal} {\bibinfo  {journal}
  {Proc. Natl. Acad. Sci. USA}\ }\textbf {\bibinfo {volume} {105}},\ \bibinfo
  {pages} {11055} (\bibinfo {year} {2008})}\BibitemShut {NoStop}%
\bibitem [{\citenamefont {Bartumeus}\ \emph {et~al.}(2002)\citenamefont
  {Bartumeus}, \citenamefont {Catalan}, \citenamefont {Fulco}, \citenamefont
  {Lyra},\ and\ \citenamefont
  {Viswanathan}}]{BartumeusCatalanFulcoLyraViswanathan:2002}%
  \BibitemOpen
  \bibfield  {author} {\bibinfo {author} {\bibfnamefont {F.}~\bibnamefont
  {Bartumeus}}, \bibinfo {author} {\bibfnamefont {J.}~\bibnamefont {Catalan}},
  \bibinfo {author} {\bibfnamefont {U.~L.}\ \bibnamefont {Fulco}}, \bibinfo
  {author} {\bibfnamefont {M.~L.}\ \bibnamefont {Lyra}},\ and\ \bibinfo
  {author} {\bibfnamefont {G.~M.}\ \bibnamefont {Viswanathan}},\ }\bibfield
  {title} {\bibinfo {title} {Optimizing the encounter rate in biological
  interactions: L\'{e}vy versus {B}rownian strategies},\ }\href@noop {}
  {\bibfield  {journal} {\bibinfo  {journal} {Phys. Rev. Lett.}\ }\textbf
  {\bibinfo {volume} {88}},\ \bibinfo {pages} {097901} (\bibinfo {year}
  {2002})}\BibitemShut {NoStop}%
\bibitem [{\citenamefont {Stone}(1975)}]{Stone:1975}%
  \BibitemOpen
  \bibfield  {author} {\bibinfo {author} {\bibfnamefont {L.~D.}\ \bibnamefont
  {Stone}},\ }\href@noop {} {\emph {\bibinfo {title} {Theory of Optimal
  Search}}}\ (\bibinfo  {publisher} {Academic Press},\ \bibinfo {address} {New
  York},\ \bibinfo {year} {1975})\BibitemShut {NoStop}%
\bibitem [{\citenamefont {Coppey}\ \emph {et~al.}(2004)\citenamefont {Coppey},
  \citenamefont {B\'{e}nichou}, \citenamefont {Voituriez},\ and\ \citenamefont
  {Moreau}}]{CoppeyBenichouVoituriezMoreau:2004}%
  \BibitemOpen
  \bibfield  {author} {\bibinfo {author} {\bibfnamefont {M.}~\bibnamefont
  {Coppey}}, \bibinfo {author} {\bibfnamefont {O.}~\bibnamefont
  {B\'{e}nichou}}, \bibinfo {author} {\bibfnamefont {R.}~\bibnamefont
  {Voituriez}},\ and\ \bibinfo {author} {\bibfnamefont {M.}~\bibnamefont
  {Moreau}},\ }\bibfield  {title} {\bibinfo {title} {Kinetics of target site
  localization of a protein on {DNA}: A stochastic approach},\ }\href@noop {}
  {\bibfield  {journal} {\bibinfo  {journal} {Biophys. J.}\ }\textbf {\bibinfo
  {volume} {87}},\ \bibinfo {pages} {1640} (\bibinfo {year}
  {2004})}\BibitemShut {NoStop}%
\bibitem [{\citenamefont {Xu}\ and\ \citenamefont
  {Deng}(2018{\natexlab{a}})}]{XuDeng:2018}%
  \BibitemOpen
  \bibfield  {author} {\bibinfo {author} {\bibfnamefont {P.~B.}\ \bibnamefont
  {Xu}}\ and\ \bibinfo {author} {\bibfnamefont {W.~H.}\ \bibnamefont {Deng}},\
  }\bibfield  {title} {\bibinfo {title} {Fractional compound poisson processes
  with multiple internal states},\ }\href@noop {} {\bibfield  {journal}
  {\bibinfo  {journal} {Math. Model. Nat. Phenom}\ }\textbf {\bibinfo {volume}
  {13}},\ \bibinfo {pages} {10} (\bibinfo {year}
  {2018}{\natexlab{a}})}\BibitemShut {NoStop}%
\bibitem [{\citenamefont {Xu}\ and\ \citenamefont
  {Deng}(2018{\natexlab{b}})}]{XuDeng:2018-2}%
  \BibitemOpen
  \bibfield  {author} {\bibinfo {author} {\bibfnamefont {P.~B.}\ \bibnamefont
  {Xu}}\ and\ \bibinfo {author} {\bibfnamefont {W.~H.}\ \bibnamefont {Deng}},\
  }\bibfield  {title} {\bibinfo {title} {L\'{e}vy walk with multiple internal
  states},\ }\href@noop {} {\bibfield  {journal} {\bibinfo  {journal} {J. Stat.
  Phys.}\ }\textbf {\bibinfo {volume} {173}},\ \bibinfo {pages} {1598}
  (\bibinfo {year} {2018}{\natexlab{b}})}\BibitemShut {NoStop}%
\bibitem [{\citenamefont {Song}\ \emph {et~al.}(2018)\citenamefont {Song},
  \citenamefont {Moon}, \citenamefont {Jeon},\ and\ \citenamefont
  {Park}}]{SongMoonJeonPark:2018}%
  \BibitemOpen
  \bibfield  {author} {\bibinfo {author} {\bibfnamefont {M.~S.}\ \bibnamefont
  {Song}}, \bibinfo {author} {\bibfnamefont {H.~C.}\ \bibnamefont {Moon}},
  \bibinfo {author} {\bibfnamefont {J.-H.}\ \bibnamefont {Jeon}},\ and\
  \bibinfo {author} {\bibfnamefont {H.~Y.}\ \bibnamefont {Park}},\ }\bibfield
  {title} {\bibinfo {title} {Neuronal messenger ribonucleoprotein transport
  follows an aging {L}\'{e}vy walk},\ }\href@noop {} {\bibfield  {journal}
  {\bibinfo  {journal} {Nat. Commun.}\ }\textbf {\bibinfo {volume} {9}},\
  \bibinfo {pages} {344} (\bibinfo {year} {2018})}\BibitemShut {NoStop}%
\bibitem [{\citenamefont {Barkai}\ and\ \citenamefont
  {Cheng}(2003)}]{BarkaiCheng:2003}%
  \BibitemOpen
  \bibfield  {author} {\bibinfo {author} {\bibfnamefont {E.}~\bibnamefont
  {Barkai}}\ and\ \bibinfo {author} {\bibfnamefont {Y.-C.}\ \bibnamefont
  {Cheng}},\ }\bibfield  {title} {\bibinfo {title} {Aging continuous time
  random walks},\ }\href@noop {} {\bibfield  {journal} {\bibinfo  {journal} {J.
  Chem. Phys.}\ }\textbf {\bibinfo {volume} {118}},\ \bibinfo {pages} {6167}
  (\bibinfo {year} {2003})}\BibitemShut {NoStop}%
\bibitem [{\citenamefont {Schulz}\ \emph {et~al.}(2014)\citenamefont {Schulz},
  \citenamefont {Barkai},\ and\ \citenamefont
  {Metzler}}]{SchulzBarkaiMetzler:2014}%
  \BibitemOpen
  \bibfield  {author} {\bibinfo {author} {\bibfnamefont {J.~H.~P.}\
  \bibnamefont {Schulz}}, \bibinfo {author} {\bibfnamefont {E.}~\bibnamefont
  {Barkai}},\ and\ \bibinfo {author} {\bibfnamefont {R.}~\bibnamefont
  {Metzler}},\ }\bibfield  {title} {\bibinfo {title} {Aging renewal theory and
  application to random walks},\ }\href@noop {} {\bibfield  {journal} {\bibinfo
   {journal} {Phys. Rev. X}\ }\textbf {\bibinfo {volume} {4}},\ \bibinfo
  {pages} {011028} (\bibinfo {year} {2014})}\BibitemShut {NoStop}%
\bibitem [{\citenamefont {Magdziarz}\ and\ \citenamefont
  {Zorawik}(2017)}]{MagdziarzZorawik:2017}%
  \BibitemOpen
  \bibfield  {author} {\bibinfo {author} {\bibfnamefont {M.}~\bibnamefont
  {Magdziarz}}\ and\ \bibinfo {author} {\bibfnamefont {T.}~\bibnamefont
  {Zorawik}},\ }\bibfield  {title} {\bibinfo {title} {Aging ballistic
  {L}\'{e}vy walks},\ }\href@noop {} {\bibfield  {journal} {\bibinfo  {journal}
  {Phys. Rev. E}\ }\textbf {\bibinfo {volume} {95}},\ \bibinfo {pages} {022126}
  (\bibinfo {year} {2017})}\BibitemShut {NoStop}%
\bibitem [{\citenamefont {Sokolov}\ \emph {et~al.}(2001)\citenamefont
  {Sokolov}, \citenamefont {Blumen},\ and\ \citenamefont
  {Klafter}}]{SokolovBlumenKlafter:2001}%
  \BibitemOpen
  \bibfield  {author} {\bibinfo {author} {\bibfnamefont {I.~M.}\ \bibnamefont
  {Sokolov}}, \bibinfo {author} {\bibfnamefont {A.}~\bibnamefont {Blumen}},\
  and\ \bibinfo {author} {\bibfnamefont {J.}~\bibnamefont {Klafter}},\
  }\bibfield  {title} {\bibinfo {title} {Linear response in complex systems:
  {CTRW} and the fractional {F}okker-{P}lanck equations},\ }\href@noop {}
  {\bibfield  {journal} {\bibinfo  {journal} {Physica A}\ }\textbf {\bibinfo
  {volume} {302}},\ \bibinfo {pages} {268} (\bibinfo {year}
  {2001})}\BibitemShut {NoStop}%
\bibitem [{\citenamefont {Allegrini}\ \emph {et~al.}(2002)\citenamefont
  {Allegrini}, \citenamefont {Bellazzini}, \citenamefont {Bramanti},
  \citenamefont {Ignaccolo}, \citenamefont {Grigolini},\ and\ \citenamefont
  {Yang}}]{Allegrini--etal:2002}%
  \BibitemOpen
  \bibfield  {author} {\bibinfo {author} {\bibfnamefont {P.}~\bibnamefont
  {Allegrini}}, \bibinfo {author} {\bibfnamefont {J.}~\bibnamefont
  {Bellazzini}}, \bibinfo {author} {\bibfnamefont {G.}~\bibnamefont
  {Bramanti}}, \bibinfo {author} {\bibfnamefont {M.}~\bibnamefont {Ignaccolo}},
  \bibinfo {author} {\bibfnamefont {P.}~\bibnamefont {Grigolini}},\ and\
  \bibinfo {author} {\bibfnamefont {J.}~\bibnamefont {Yang}},\ }\bibfield
  {title} {\bibinfo {title} {Scaling breakdown: A signature of aging},\
  }\href@noop {} {\bibfield  {journal} {\bibinfo  {journal} {Phys. Rev. E}\
  }\textbf {\bibinfo {volume} {66}},\ \bibinfo {pages} {015101(R)} (\bibinfo
  {year} {2002})}\BibitemShut {NoStop}%
\bibitem [{\citenamefont {Montroll}\ and\ \citenamefont
  {Weiss}(1965)}]{MontrollWeiss:1965}%
  \BibitemOpen
  \bibfield  {author} {\bibinfo {author} {\bibfnamefont {E.~W.}\ \bibnamefont
  {Montroll}}\ and\ \bibinfo {author} {\bibfnamefont {G.~H.}\ \bibnamefont
  {Weiss}},\ }\bibfield  {title} {\bibinfo {title} {Random walks on lattices.
  {II}},\ }\href@noop {} {\bibfield  {journal} {\bibinfo  {journal} {J. Math.
  Phys.}\ }\textbf {\bibinfo {volume} {6}},\ \bibinfo {pages} {167} (\bibinfo
  {year} {1965})}\BibitemShut {NoStop}%
\bibitem [{\citenamefont {Zaburdaev}\ \emph {et~al.}(2015)\citenamefont
  {Zaburdaev}, \citenamefont {Denisov},\ and\ \citenamefont
  {Klafter}}]{ZaburdaevDenisovKlafter:2015}%
  \BibitemOpen
  \bibfield  {author} {\bibinfo {author} {\bibfnamefont {V.}~\bibnamefont
  {Zaburdaev}}, \bibinfo {author} {\bibfnamefont {S.}~\bibnamefont {Denisov}},\
  and\ \bibinfo {author} {\bibfnamefont {J.}~\bibnamefont {Klafter}},\
  }\bibfield  {title} {\bibinfo {title} {L\'{e}vy walks},\ }\href@noop {}
  {\bibfield  {journal} {\bibinfo  {journal} {Rev. Mod. Phys.}\ }\textbf
  {\bibinfo {volume} {87}},\ \bibinfo {pages} {483} (\bibinfo {year}
  {2015})}\BibitemShut {NoStop}%
\bibitem [{\citenamefont {Shlesinger}\ \emph {et~al.}(1995)\citenamefont
  {Shlesinger}, \citenamefont {Zaslavsky},\ and\ \citenamefont
  {Frisch}}]{ShlesingerZaslavskyFrisch:1995}%
  \BibitemOpen
  \bibfield  {author} {\bibinfo {author} {\bibfnamefont {M.~F.}\ \bibnamefont
  {Shlesinger}}, \bibinfo {author} {\bibfnamefont {G.~M.}\ \bibnamefont
  {Zaslavsky}},\ and\ \bibinfo {author} {\bibfnamefont {U.}~\bibnamefont
  {Frisch}},\ }\href@noop {} {\emph {\bibinfo {title} {L\'{e}vy Flights and
  Related Topics}}}\ (\bibinfo  {publisher} {Springer-Verlag},\ \bibinfo
  {address} {Berlin},\ \bibinfo {year} {1995})\BibitemShut {NoStop}%
\bibitem [{\citenamefont {Metzler}\ and\ \citenamefont
  {Klafter}(2000)}]{MetzlerKlafter:2000}%
  \BibitemOpen
  \bibfield  {author} {\bibinfo {author} {\bibfnamefont {R.}~\bibnamefont
  {Metzler}}\ and\ \bibinfo {author} {\bibfnamefont {J.}~\bibnamefont
  {Klafter}},\ }\bibfield  {title} {\bibinfo {title} {The random walk's guide
  to anomalous diffusion: a fractional dynamics approach},\ }\href@noop {}
  {\bibfield  {journal} {\bibinfo  {journal} {Phys. Rep.}\ }\textbf {\bibinfo
  {volume} {339}},\ \bibinfo {pages} {1} (\bibinfo {year} {2000})}\BibitemShut
  {NoStop}%
\bibitem [{\citenamefont {Godr{\`{e}}che}\ and\ \citenamefont
  {Luck}(2001)}]{GodrecheLuck:2001}%
  \BibitemOpen
  \bibfield  {author} {\bibinfo {author} {\bibfnamefont {C.}~\bibnamefont
  {Godr{\`{e}}che}}\ and\ \bibinfo {author} {\bibfnamefont {J.~M.}\
  \bibnamefont {Luck}},\ }\bibfield  {title} {\bibinfo {title} {Statistics of
  the occupation time of renewal processes},\ }\href@noop {} {\bibfield
  {journal} {\bibinfo  {journal} {J. Stat. Phys.}\ }\textbf {\bibinfo {volume}
  {104}},\ \bibinfo {pages} {489} (\bibinfo {year} {2001})}\BibitemShut
  {NoStop}%
\bibitem [{\citenamefont {Dechant}\ \emph {et~al.}(2014)\citenamefont
  {Dechant}, \citenamefont {Lutz}, \citenamefont {Kessler},\ and\ \citenamefont
  {Barkai}}]{DechantLutzKesslerBarkai:2014}%
  \BibitemOpen
  \bibfield  {author} {\bibinfo {author} {\bibfnamefont {A.}~\bibnamefont
  {Dechant}}, \bibinfo {author} {\bibfnamefont {E.}~\bibnamefont {Lutz}},
  \bibinfo {author} {\bibfnamefont {D.~A.}\ \bibnamefont {Kessler}},\ and\
  \bibinfo {author} {\bibfnamefont {E.}~\bibnamefont {Barkai}},\ }\bibfield
  {title} {\bibinfo {title} {Scaling {G}reen-{K}ubo relation and application to
  three aging systems},\ }\href@noop {} {\bibfield  {journal} {\bibinfo
  {journal} {Phys. Rev. X}\ }\textbf {\bibinfo {volume} {4}},\ \bibinfo {pages}
  {011022} (\bibinfo {year} {2014})}\BibitemShut {NoStop}%
\bibitem [{\citenamefont {Meyer}\ \emph {et~al.}(2017)\citenamefont {Meyer},
  \citenamefont {Barkai},\ and\ \citenamefont {Kantz}}]{MeyerBarkaiKantz:2017}%
  \BibitemOpen
  \bibfield  {author} {\bibinfo {author} {\bibfnamefont {P.}~\bibnamefont
  {Meyer}}, \bibinfo {author} {\bibfnamefont {E.}~\bibnamefont {Barkai}},\ and\
  \bibinfo {author} {\bibfnamefont {H.}~\bibnamefont {Kantz}},\ }\bibfield
  {title} {\bibinfo {title} {Scale-invariant {G}reen-{K}ubo relation for
  time-averaged diffusivity},\ }\href@noop {} {\bibfield  {journal} {\bibinfo
  {journal} {Phys. Rev. E}\ }\textbf {\bibinfo {volume} {96}},\ \bibinfo
  {pages} {062122} (\bibinfo {year} {2017})}\BibitemShut {NoStop}%
\bibitem [{\citenamefont {Metzler}\ \emph {et~al.}(2014)\citenamefont
  {Metzler}, \citenamefont {Jeon}, \citenamefont {Cherstvy},\ and\
  \citenamefont {Barkai}}]{MetzlerJeonCherstvyBarkai:2014}%
  \BibitemOpen
  \bibfield  {author} {\bibinfo {author} {\bibfnamefont {R.}~\bibnamefont
  {Metzler}}, \bibinfo {author} {\bibfnamefont {J.-H.}\ \bibnamefont {Jeon}},
  \bibinfo {author} {\bibfnamefont {A.~G.}\ \bibnamefont {Cherstvy}},\ and\
  \bibinfo {author} {\bibfnamefont {E.}~\bibnamefont {Barkai}},\ }\bibfield
  {title} {\bibinfo {title} {Anomalous diffusion models and their properties:
  non-stationarity, non-ergodicity, and ageing at the centenary of single
  particle tracking},\ }\href@noop {} {\bibfield  {journal} {\bibinfo
  {journal} {Phys. Chem. Chem. Phys.}\ }\textbf {\bibinfo {volume} {16}},\
  \bibinfo {pages} {24128} (\bibinfo {year} {2014})}\BibitemShut {NoStop}%
\bibitem [{\citenamefont {Godec}\ and\ \citenamefont
  {Metzler}(2001)}]{GodecMetzler:2013}%
  \BibitemOpen
  \bibfield  {author} {\bibinfo {author} {\bibfnamefont {A.}~\bibnamefont
  {Godec}}\ and\ \bibinfo {author} {\bibfnamefont {R.}~\bibnamefont
  {Metzler}},\ }\bibfield  {title} {\bibinfo {title} {Finite-time effects and
  ultraweak ergodicity breaking in superdiffusive dynamics},\ }\href@noop {}
  {\bibfield  {journal} {\bibinfo  {journal} {Phys. Rev. Lett.}\ }\textbf
  {\bibinfo {volume} {104}},\ \bibinfo {pages} {489} (\bibinfo {year}
  {2001})}\BibitemShut {NoStop}%
\bibitem [{\citenamefont {Froemberg}\ and\ \citenamefont
  {Barkai}(2013)}]{FroembergBarkai:2013}%
  \BibitemOpen
  \bibfield  {author} {\bibinfo {author} {\bibfnamefont {D.}~\bibnamefont
  {Froemberg}}\ and\ \bibinfo {author} {\bibfnamefont {E.}~\bibnamefont
  {Barkai}},\ }\bibfield  {title} {\bibinfo {title} {Time-averaged {E}instein
  relation and fluctuating diffusivities for the {L}\'{e}vy walk},\ }\href@noop
  {} {\bibfield  {journal} {\bibinfo  {journal} {Phys. Rev. E}\ }\textbf
  {\bibinfo {volume} {87}},\ \bibinfo {pages} {030104(R)} (\bibinfo {year}
  {2013})}\BibitemShut {NoStop}%
\bibitem [{\citenamefont {Solomon}\ \emph {et~al.}(1993)\citenamefont
  {Solomon}, \citenamefont {Weeks},\ and\ \citenamefont
  {Swinney}}]{SolomonWeeksSwinney:1993}%
  \BibitemOpen
  \bibfield  {author} {\bibinfo {author} {\bibfnamefont {T.~H.}\ \bibnamefont
  {Solomon}}, \bibinfo {author} {\bibfnamefont {E.~R.}\ \bibnamefont {Weeks}},\
  and\ \bibinfo {author} {\bibfnamefont {H.~L.}\ \bibnamefont {Swinney}},\
  }\bibfield  {title} {\bibinfo {title} {Observation of anomalous diffusion and
  {L}\'{e}vy flights in a 2-dimensional rotating flow},\ }\href@noop {}
  {\bibfield  {journal} {\bibinfo  {journal} {Phys. Rev. Lett.}\ }\textbf
  {\bibinfo {volume} {71}},\ \bibinfo {pages} {3975} (\bibinfo {year}
  {1993})}\BibitemShut {NoStop}%
\bibitem [{\citenamefont {Klafter}\ and\ \citenamefont
  {Zumofen}(1994)}]{KlafterZumofen:1994}%
  \BibitemOpen
  \bibfield  {author} {\bibinfo {author} {\bibfnamefont {J.}~\bibnamefont
  {Klafter}}\ and\ \bibinfo {author} {\bibfnamefont {G.}~\bibnamefont
  {Zumofen}},\ }\bibfield  {title} {\bibinfo {title} {{L}\'{e}vy statistics in
  a {H}amiltonian system},\ }\href@noop {} {\bibfield  {journal} {\bibinfo
  {journal} {Phys. Rev. E}\ }\textbf {\bibinfo {volume} {49}},\ \bibinfo
  {pages} {4873} (\bibinfo {year} {1994})}\BibitemShut {NoStop}%
\bibitem [{\citenamefont {Cox}(1962)}]{Cox:1962}%
  \BibitemOpen
  \bibfield  {author} {\bibinfo {author} {\bibfnamefont {D.~R.}\ \bibnamefont
  {Cox}},\ }\href@noop {} {\emph {\bibinfo {title} {Renewal Theory}}}\
  (\bibinfo  {publisher} {Methuen},\ \bibinfo {address} {London},\ \bibinfo
  {year} {1962})\BibitemShut {NoStop}%
\bibitem [{\citenamefont {Klafter}\ and\ \citenamefont
  {Zumofen}(1993)}]{KlafterZumofen:1993}%
  \BibitemOpen
  \bibfield  {author} {\bibinfo {author} {\bibfnamefont {J.}~\bibnamefont
  {Klafter}}\ and\ \bibinfo {author} {\bibfnamefont {G.}~\bibnamefont
  {Zumofen}},\ }\bibfield  {title} {\bibinfo {title} {Dynamically generated
  enhanced diffusion: the stationary state case},\ }\href@noop {} {\bibfield
  {journal} {\bibinfo  {journal} {Physica A}\ }\textbf {\bibinfo {volume}
  {196}},\ \bibinfo {pages} {102} (\bibinfo {year} {1993})}\BibitemShut
  {NoStop}%
\bibitem [{\citenamefont {Wang}\ \emph {et~al.}(2019)\citenamefont {Wang},
  \citenamefont {Chen},\ and\ \citenamefont {Deng}}]{WangChenDeng:2019}%
  \BibitemOpen
  \bibfield  {author} {\bibinfo {author} {\bibfnamefont {X.~D.}\ \bibnamefont
  {Wang}}, \bibinfo {author} {\bibfnamefont {Y.}~\bibnamefont {Chen}},\ and\
  \bibinfo {author} {\bibfnamefont {W.~H.}\ \bibnamefont {Deng}},\ }\bibfield
  {title} {\bibinfo {title} {L\'{e}vy-walk-like {L}angevin dynamics},\
  }\href@noop {} {\bibfield  {journal} {\bibinfo  {journal} {New J. Phys.}\
  }\textbf {\bibinfo {volume} {21}},\ \bibinfo {pages} {013024} (\bibinfo
  {year} {2019})}\BibitemShut {NoStop}%
\bibitem [{\citenamefont {Miyaguchi}\ \emph {et~al.}(2016)\citenamefont
  {Miyaguchi}, \citenamefont {Akimoto},\ and\ \citenamefont
  {Yamamoto}}]{MiyaguchiAkimotoYamamoto:2016}%
  \BibitemOpen
  \bibfield  {author} {\bibinfo {author} {\bibfnamefont {T.}~\bibnamefont
  {Miyaguchi}}, \bibinfo {author} {\bibfnamefont {T.}~\bibnamefont {Akimoto}},\
  and\ \bibinfo {author} {\bibfnamefont {E.}~\bibnamefont {Yamamoto}},\
  }\bibfield  {title} {\bibinfo {title} {Langevin equation with fluctuating
  diffusivity: A two-state model},\ }\href@noop {} {\bibfield  {journal}
  {\bibinfo  {journal} {Phys. Rev. E}\ }\textbf {\bibinfo {volume} {94}},\
  \bibinfo {pages} {012109} (\bibinfo {year} {2016})}\BibitemShut {NoStop}%
\bibitem [{\citenamefont {Golding}\ and\ \citenamefont
  {Cox}(2006)}]{GoldingCox:2006}%
  \BibitemOpen
  \bibfield  {author} {\bibinfo {author} {\bibfnamefont {I.}~\bibnamefont
  {Golding}}\ and\ \bibinfo {author} {\bibfnamefont {E.~C.}\ \bibnamefont
  {Cox}},\ }\bibfield  {title} {\bibinfo {title} {Physical nature of bacterial
  cytoplasm},\ }\href@noop {} {\bibfield  {journal} {\bibinfo  {journal} {Phys.
  Rev. Lett.}\ }\textbf {\bibinfo {volume} {96}},\ \bibinfo {pages} {098102}
  (\bibinfo {year} {2006})}\BibitemShut {NoStop}%
\bibitem [{\citenamefont {Chen}\ \emph {et~al.}(2019)\citenamefont {Chen},
  \citenamefont {Wang},\ and\ \citenamefont {Deng}}]{ChenWangDeng:2019}%
  \BibitemOpen
  \bibfield  {author} {\bibinfo {author} {\bibfnamefont {Y.}~\bibnamefont
  {Chen}}, \bibinfo {author} {\bibfnamefont {X.~D.}\ \bibnamefont {Wang}},\
  and\ \bibinfo {author} {\bibfnamefont {W.~H.}\ \bibnamefont {Deng}},\
  }\bibfield  {title} {\bibinfo {title} {Langevin dynamics for a {L}\'{e}vy
  walk with memory},\ }\href@noop {} {\bibfield  {journal} {\bibinfo  {journal}
  {Phys. Rev. E}\ }\textbf {\bibinfo {volume} {99}},\ \bibinfo {pages} {012135}
  (\bibinfo {year} {2019})}\BibitemShut {NoStop}%
\bibitem [{\citenamefont {Kami\'{n}ska}\ and\ \citenamefont
  {Srokowski}(2018)}]{KaminskaSrokowski:2018}%
  \BibitemOpen
  \bibfield  {author} {\bibinfo {author} {\bibfnamefont {A.}~\bibnamefont
  {Kami\'{n}ska}}\ and\ \bibinfo {author} {\bibfnamefont {T.}~\bibnamefont
  {Srokowski}},\ }\bibfield  {title} {\bibinfo {title} {L\'{e}vy walks with
  variable waiting time: A ballistic case},\ }\href@noop {} {\bibfield
  {journal} {\bibinfo  {journal} {Phys. Rev. E}\ }\textbf {\bibinfo {volume}
  {97}},\ \bibinfo {pages} {062120} (\bibinfo {year} {2018})}\BibitemShut
  {NoStop}%
\end{thebibliography}%

\end{document}